# Effect of substituting non-polar chains with polar chains on the structural dynamics of small organic molecule and polymer semiconductors

Anne A. Y. Guilbert,*[a] Zachary S. Parr,[b] Theo Kreouzis,[c] Duncan J. Woods,[d] Reiner S. Sprick,[d] Isaac Abrahams,[b] Christian B. Nielsen[b] and Mohamed Zbiri*[e]

[a]Department of Physics and Centre for Plastic Electronics, Imperial College London, Prince Consort Road, London SW7 2AZ, United Kingdom.
[b]Materials Research Institute and School of Biological and Chemical Sciences, Queen Mary University of London, Mile End Road, London E1 4NS, United Kingdom.
[c]School of Physics and Astronomy, Queen Mary University of London, London E1 4NS United Kingdom
[d]Department of Chemistry and Material Innovation Factory, University of Liverpool, Crown Street, Liverpool L69 7ZD, United Kingdom.
[e]Institut Laue-Langevin, 71 Avenue des Martyrs, Grenoble Cedex 9 38042, France.

*a.guilbert09@imperial.ac.uk
*zbiri@ill.fr

The processability and optoelectronic properties of organic semiconductors can be tuned and manipulated via chemical design. The substitution of the popular alkyl side chains by oligoethers has recently been successful for applications such as bioelectronic sensors and photocatalytic water-splitting. Beyond the differences in polarity, the carbon-oxygen bond in oligoethers is likely to render the system softer and more prone to dynamical disorder that can be detrimental to charge transport for example. In this context, we use neutron spectroscopy as a master method of probe, in addition to characterisation techniques such as X-Ray diffraction, differential scanning calorimetry and polarized optical microscopy to study the effect of the substitution of n-hexyl (Hex) chains by triethylene glycol (TEG) chains on the structural dynamics of two organic semiconducting materials: a phenylene-bithiophene-phenylene (PTTP) small molecule and a fluorene-co-dibenzothiophene (FS) polymer. Counterintuitively, inelastic neutron scattering (INS) reveals a general softening of the modes of PTTP and FS materials with Hex chains, pointing towards an increased dynamical disorder in the Hex-based systems. However, temperature-dependent X-Ray and neutron diffraction as well as INS and differential scanning calorimetry evidence an extra reversible transition close to room temperature for PTTP with TEG chains. The observed extra structural transition, which is not accompanied by a change in birefringence, can also be observed by quasi-elastic neutron scattering (QENS). A fastening of the TEG chains dynamics is observed in the case of PTTP and not FS. We therefore assign this transition to the melt of the TEG chains. Overall the TEG chains are promoting dynamical order at room temperature, but if crystallising, may introduce an extra reversible structural transition above room temperature leading to thermal instabilities. Ultimately, a deeper understanding of chain polarity and structural dynamics can help guide new materials design and navigate the intricate balance between electronic charge transport and aqueous swelling that is being sought for a number of emerging organic electronic and bioelectronic applications.

## Introduction

Chemical design is a ubiquitous aspect of organic electronics, which has helped advance mature technologies such as organic light-emitting displays to commercialisation and allowed for other technologies such as organic photovoltaics to undergo rapid performance improvements in recent years.[1] Chemical design is moreover pushing new emerging technologies to the forefront of organic electronics research with promising developments in areas such as organic bioelectronics and photocatalysis.[2–5]

Being able to understand and manipulate electrical charge transport across several metal-organic and organic-organic interfaces in these electronic devices has necessitated judicious control of frontier energy levels in organic semiconductor materials. This is now a well-established area with many clear structure-property relations and numerous powerful chemical design tools such as push-pull (donor-acceptor) type molecular structures, non-covalent conformational locking interactions and various heteroatom effects.[1,6]

A significant selling point of organic electronics is the solution processability of the organic semiconductor materials, which is typically imparted by alkyl chains covalently attached to the π-conjugated backbone. While early semiconductor materials such as polythiophenes and polyfluorenes were solubilised with relatively short and linear alkyl chains such as n-hexyl and n-octyl chains, the subsequent development of new semiconductors with fused aromatic cores and lower alkyl chain densities brought along longer linear chains (e.g. n-tetradecyl and n-hexadecyl) as well as a large family of branched alkyl chains.[7] Moving away from purely alkyl based chains, solubilising chains with for instance siloxane and amide functionalities have been used to improve solid-state structural organisation, charge transport and self-healing properties.[8,9]

The introduction of highly polar chains, in particular those based on oligoethers, has more recently emerged as a particularly powerful tool to tune a variety of materials properties such as the structural packing, dielectric constant, water swelling and interactions with polar and ionic species including molecular dopants and biologically relevant ions and molecules.[10] In several instances, π-conjugated polymers with oligoether side chains have been found to shorten the intermolecular π-stacking distance between polymer chains compared to their alkylated counterparts.[11,12] Stronger π-stacking interactions in these glycolated materials have led to higher degrees of crystallinity and better charge transport in field-effect transistors. Organic thermoelectrics rely on effective p- and n-type doping to increase the electrical conductivity of hole and electron transporting organic semiconductors which in turn allow for energy harvesting across a temperature gradient. A morphologically stable dispersion of molecular dopants within the organic semiconductor at elevated temperatures is thus an important prerequisite in a thermoelectric generator. In this context, glycolated semiconductors have been shown to facilitate higher electrical conductivity and better thermal stability than their alkylated counterparts. This has been observed for several materials classes including thiophene based p-type polymers, naphthalene diimide based n-type polymers and molecular fullerenes.[13–15] In the case of n-type doping of a glycolated fullerene derivative, the increased molecular polarity of the fullerene (compared to an alkylated analogue) is hypothesised to improve the miscibility of the fullerene and the molecular dopant and also solubilise the cationic doping product to a greater extent. Glycolated materials have likewise made a significant impact on the emerging field of organic bioelectronics and in particular on the development of new active materials for aqueous electrolyte-gated transistors which are promising devices for a variety of biosensing applications.[16,17] While alkylated semiconductors due to their hydrophobic nature only allow for interfacial charge transport in a transistor, the glycolated counterparts can facilitate penetration of ions into the bulk of the semiconductor and hence shift the mode of operation to bulk charge transport.[18] During recent years, this has been documented for thiophene and isoindigo based p-type polymers, naphthalene diimide based n-type polymers as well as both p- and n-type molecular systems.[16,17,19–21] Oligoether side chains have been recently introduced for photocatalytic water-splitting applications. In this context, the polar side chains are thought to improve the photocatalytic activity by leading to polymer swelling in water (higher surface area for the catalytic reaction to take place) and longer-lived electrons due to a change in the dielectric constant of the environment and thus, stabilisation of the charges.[22,23] It is also found to interact strongly with palladium co-catalyst.[22]

While it is clear that oligoether chains present a relatively new and highly valuable addition to the organic chemistry toolbox used for designing and synthesising organic electronic materials, the fundamental differences between oligoethers and alkyl chains in the context of organic electronics have not been studied in much detail across several systems. The carbon-oxygen bond in oligoethers is likely to rotate more freely than carbon-carbon bonds in alkyl chains, leading to a more flexible system. As discussed above, this can lead to stronger intermolecular interactions. On the other hand, the dynamical disorder can also influence charge transport negatively.[24,25]

Motivated by this increased attention and significant promise of glycolated chains as a new chemical design tool in the wider field of organic electronics, we compare, herein, two organic semiconductor materials, a phenylene-bithiophene-phenylene (PTTP) molecule[20,26,27] and a fluorene-co-dibenzothiophene sulfone polymer (FS), with each organic semiconductor bearing either nonpolar n-hexyl (Hex) or polar triethylene glycol (TEG) based solubilising chains as illustrated in Figure 1.[23]

Attachment of solubilising chains onto the sp3-hybridised bridgehead atom in the fluorene copolymer is a common functionalisation strategy, not only for light-emitting fluorene-based materials such as PFO and F8BT but also for example for PCPDTBT and IDTBT, which are high-performing photovoltaic and transistor materials, respectively. Similarly, the peripheral linear chains on PTTP is a well-known motif seen for instance in other high-performing small molecule transistor materials such as C8-BTBT, C10-DNTT, DH-6T, and DH-FTTF. As such, we believe that the two materials studied herein represent two broad classes of organic semiconductor materials that have found widespread use across several organic electronic device applications. PTTP and FS will thus serve as exemplars for which we will study the effect of substituting the archetypical solubilising alkyl chains with polar glycolated chains on structure, morphology and dynamics of the materials. Our materials choices will moreover allow us to directly compare the impact of introducing glycolated chains onto two radically different semiconductor architectures, a relatively short molecular semiconductor, PTTP, with solubilising chains parallel to the long molecular axis tethered via sp2 carbons and a π-conjugated polymer, FS, with solubilising chains perpendicular to the polymer backbone attached on sp3 hybridised bridgeheads.

## Results and discussion

Fluorene-co-dibenzo[b,d]thiophene sulfone polymers FS-Hex and FS-TEG (Figure 1) were synthesised by Suzuki–Miyaura-type polycondensation and obtained with weight-average molecular weights of 8.2 kDa and 11.5 kDa, respectively, as previously reported.[23] The 5,5'-biphenyl-2,2'-bithiophene compounds PTTP-Hex and PTTP-TEG (Figure 1) were synthesised by Stille-coupling between the appropriately functionalised bromobenzenes and distannylated bithiophene as previously reported.[20,27] Thermogravimetric analysis reveals that all four materials are thermally stable until 300 ˚C.

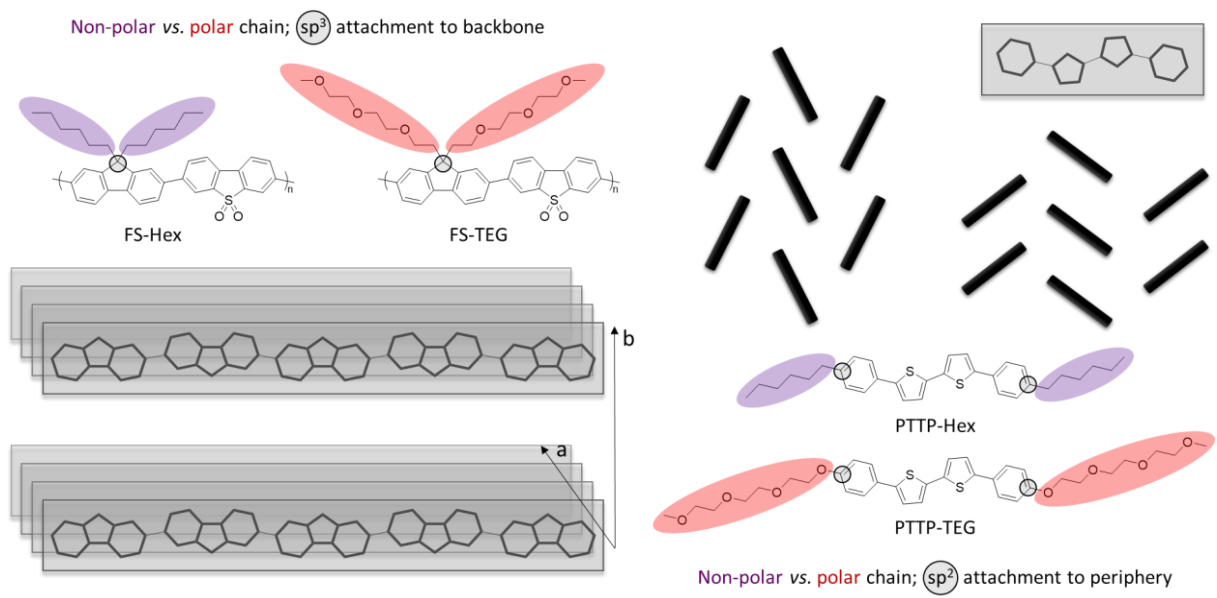

Figure 1. (Left) Alkylated (non-polar, labelled as Hex) and glycolated (polar, labelled as TEG) fluorene-co-dibenzothiophene sulfone (FS) polymer structures with side chains tethered through sp3-hybridised bridgeheads and typical polymer packing motif showing π-stacking (a) and lamellar stacking (b); (right) examples of typical molecular packing arrangements and structures of alkylated (non-polar, labelled as Hex) and glycolated (polar, labelled as TEG) phenylene-bithiophene-phenylene (PTTP) semiconductor structures with peripheral chains appended through sp2-hybridised carbon atoms.

Figure 2 (a-b) shows the neutron and (specular) X-ray diffractograms for FS-Hex, FS-TEG, PTTP-Hex, and PTTP-TEG at room temperature. While neutrons are as a bulk probe penetrate the samples, specular X-Ray diffraction provides further information regarding the out-of-plane crystallinity and orientation of the crystals with respect to the thin film substrate. X-ray and neutrons are sensitive to different elements present in the molecules and thus, provide complementary information and a more global picture of the crystal structure, especially for compounds with long alkyl chains. On the neutron diffractograms, FS-Hex exhibits only two Bragg peaks at 3.6˚ and 8.5˚, respectively. While the Bragg peak at 3.6˚ is shifted towards lower angles and appears more intense for FS-TEG, the Bragg peak at 8.5˚ is present at the same angle and no significant changes in intensity are observed for FS-TEG. FS-TEG presents additional Bragg peaks at 4.3˚, 10.5˚ and 19.5˚ and an additional broad peak at 16˚. The Bragg peaks at low Q value observed by neutron diffraction are related to the lamellar stacking of the polymers while the broad Bragg peaks at about 20˚ observed by X-ray diffraction are related to the π-π stacking of the polymer. The TEG side chains promote ordering of the π-π stacking and enhance the ordering of the lamellar stacking as observed by neutron diffraction, presumably because the side chains are more flexible and present less steric hindrance due to the reduced number of hydrogens. The diffractograms of PTTP-Hex and PTTP-TEG are also different. A Bragg peak at low Q is observed both by X-ray and neutron diffractions for PTTP-Hex. On the X-ray diffractograms, the low-Q Bragg peak is observed at 3.4 ˚ with a shoulder at 2.9 ˚. Additional Bragg peaks at 16 ˚, 16.6 ˚, 17.8 ˚, 22.5 ˚ and 25 ˚, as well as a broad peak around 19˚ are further observed by neutron diffraction for PTTP-Hex. The Bragg peaks at low angles are shifted to higher angle values for PTTP-TEG. The Bragg peaks present on the neutron diffractograms at 16 ˚, 16.6 ˚ and 17.8 ˚ are at the same Q values for PTTP-TEG but their intensities are different with respect to PTTP-Hex. On the neutron diffractogram, the broad peak around 19˚ is more resolved for PTTP-TEG with two peaks at 18.7 ˚ and 19.5 ˚. The peak at 22.5˚ is not present for PTTP-TEG. However, PTTP-TEG exhibits additional peaks at 4.4 ˚, 6 ˚, 24.2 ˚ and 26.3 ˚ as well as a broad peak around 10.3˚. Furthermore, PTTP-TEG exhibits an extra Bragg peak observed by X-ray diffraction at 7.6 ˚. The only Bragg peak presents on the specular X-ray diffractograms for PTTP-Hex corresponds to the 100 reflection according to the crystal structure in reference [28]. This suggests that the crystals of PTTP-Hex are oriented with the long axis of the unit cell perpendicular to the substrate. The shorter repeat distance for PTTP-TEG (assuming a similar perpendicular orientation and similar type unit cell) could indicate that the longer TEG chain is folded somehow while the shorter alkyl chain is fully extended. To further probe the PTTP structural order, a single crystal of PTTP-TEG was obtained for comparison with the reported PTTP-Hex single crystal. The single crystal structure of PTTP-TEG, obtained from X-Ray diffraction carried out on a single crystal grown from slow cooling of a hot toluene solution, is depicted in Figure 3 together with that of PTTP-Hex.[28] Both compounds crystallise in the monoclinic P21/c space group and the folded TEG chain hypothesised above is confirmed with the shorter long axis (a = 22.97 Å versus 25.24 Å) observed for PTTP-TEG. Both compounds pack in a typical herringbone fashion, but a lower degree of backbone planarity is observed for PTTP-TEG with phenylene-thienylene dihedral angles around 17° for PTTP-TEG compared to ~4° for PTTP-Hex.

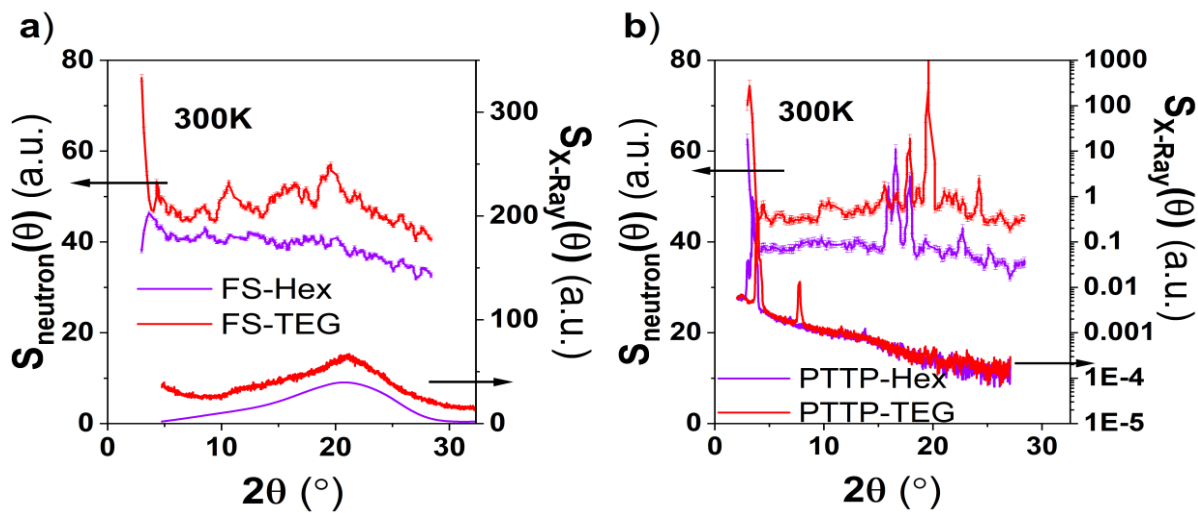

Figure 2. Comparison between the neutron and specular X-ray diffractograms of the two (a) FSs and (b) PTTPs at 300K.

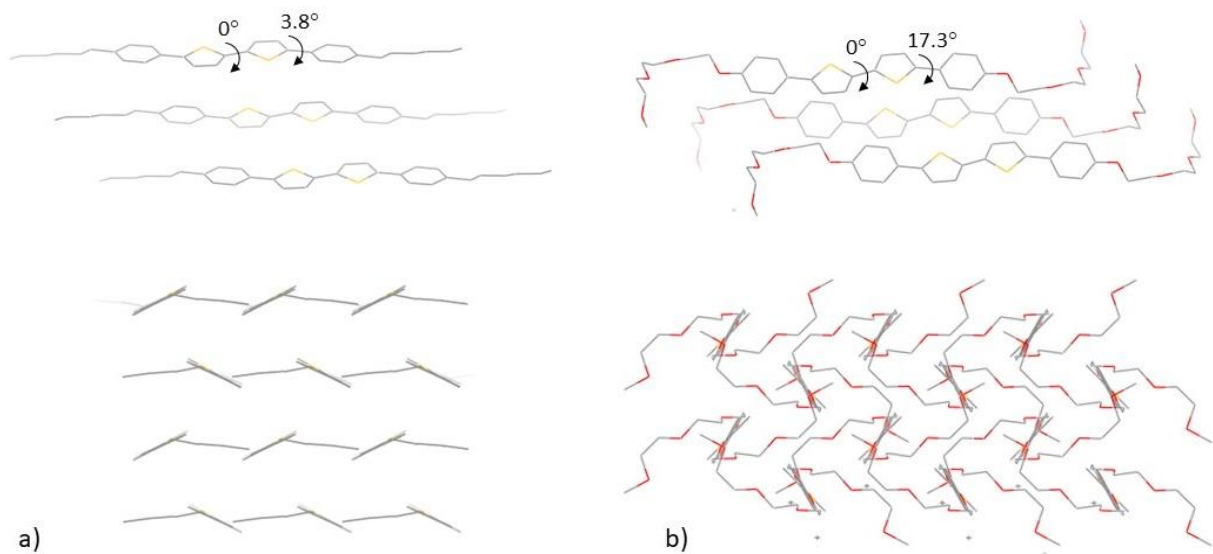

Figure 3. Single crystal structures of (a) PTTP-Hex (from ref. 28) and (b) PTTP-TEG, showing packing along the b-axis (top) and along the long molecular axis (bottom). Both compounds crystallise in the monoclinic P21/c space group with cell parameters a 25.24(3)Å b 5.666(7)Å c 9.274(12)Å, α 90° β 98.36(2)° γ 90° for PTTP-Hex and a 22.969(1)Å b 8.1561(3)Å c 8.6458(3)Å, α 90° β 100.575(4)° γ 90° for PTTP-TEG.

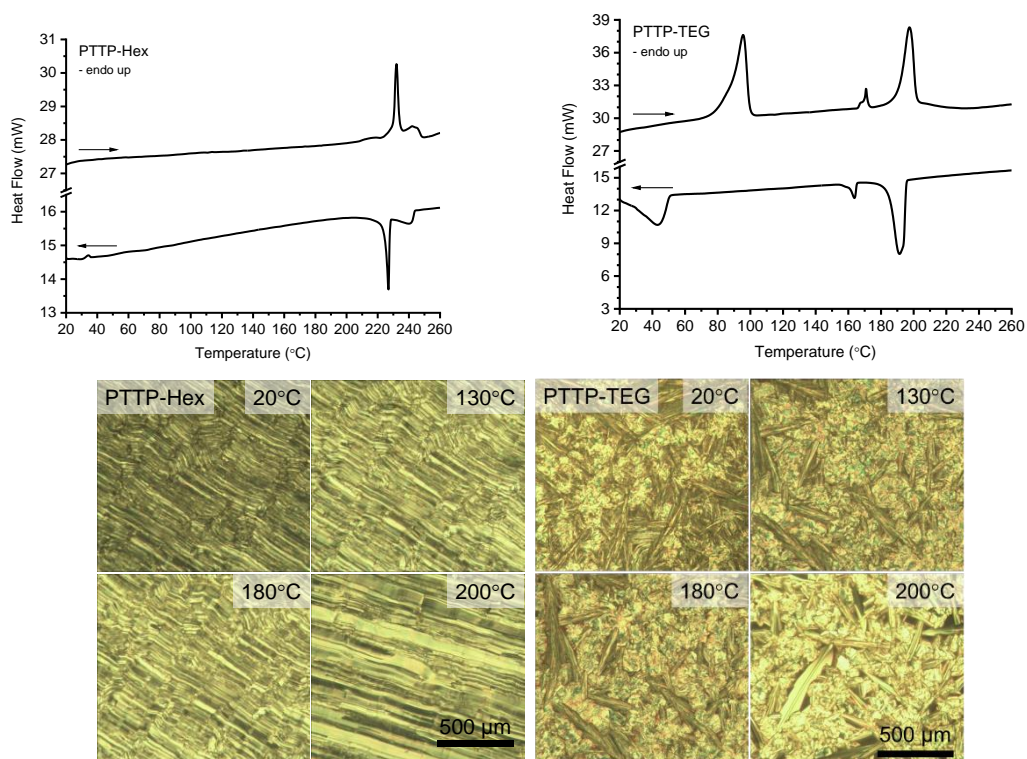

Figure 4. Differential scanning calorimetry traces (top) and polarised optical micrographs (bottom) for PTTP-Hex (left) and PTTP-TEG (right). Differential scanning calorimetry performed at a scan rate of 10 °C/min with endothermic events "up"; second heating/cooling cycle shown in both cases to avoid thermal history effects. Polarised optical micrographs recorded during the second heating cycle.

Differential scanning calorimetry (DSC) supported by polarised optical microscopy (POM) was used to probe the thermal behaviour of the two crystalline PTTP molecules as illustrated in Figure 4. Upon heating, a transition to isotropic melt is observed at 232 °C and 198 °C for PTTP-Hex and PTTP-TEG, respectively. For PTTP-TEG, two additional endothermic events at 95 °C and 171 °C are observed during the heating cycle, with associated exothermic events during the cooling cycle. Note that the transition at 171 °C presents little undercooling in contrast to the transition at 95 °C. We assume these transitions to be smectic to smectic transitions with the first transition (95 °C) most likely being a TEG chain melt and the second (171 °C) a change of orientation within the lamellae. The optical micrographs reveal different microstructure for PTTP-Hex and PTTP-TEG. PTTP-Hex is homogeneously packed, showing clear smectic homeotropic behaviour in agreement with previous literature,[29,30] while PTTP-TEG is inhomogeneous, with isotropic small crystals and larger crystals growing along a preferred orientation. We observe very little change in the birefringence by POM for PTTP-TEG across the 20 °C-200 °C temperature range indicating that liquid crystalline phases are unlikely to exist although we cannot rule out their existence entirely. In order to get further insight in the transition around 95 °C, we carried out further temperature-dependent X-Ray and neutron diffraction measurements for both sets of samples (Figure 5).

No changes are observed for both FS polymers within the measured temperature range (10-400K) (Figure 5 a-b). Some small changes in the diffractograms of PTTP-Hex at 300K before and after heating can be observed in Figure 5 c and e. Some additional Bragg peaks are present before annealing at 16°, 22.5 ° and 25 °as well as the broad peak around 19 ° in Figure 5 c. These extra peaks melt at 400K, except for the broad peak at 19 ° that sharpens upon heating, and do not reappear when cooling. The Bragg peak at 19 ° shifts to higher angles upon cooling. This points towards the presence of a second less stable polymorph at 300K depending on processing history. When heated at 400K, the low-angle peak and the Bragg peak at 17.8° (Figure 5 d) decrease in intensity while the intensity of the peaks at 16.6° increases (Figure 5 c). Heating the PTTP-Hex likely strengthens the interactions at short lattice distances at the expense of weakening them along the long axis of the unit cell. For PTTP-TEG, unlike for PTTP-Hex, no differences in the diffractograms are observed before and after heating (Figure 5 d). This can be explained by a difference in flexibility of the TEG chains as well as an increase of solubility in most common solvent leading to increased rearrangement of the molecules during processing resulting in the appearance of the most thermodynamically stable polymorph for PTTP-TEG. The diffractogram at 200K are similar to the one at 300 K and small changes can be observed at 10 K (Figure 5 d). The peaks at 16 °, 17.8 ° and 24.2 ° are slightly shifted towards higher angles and the intensity of the peaks at 17.8 °, 18.7 ° and 19.5 ° are increased. Those observations are expected from normal compression due to cooling. Interestingly, large differences in both neutron and X-ray diffractograms can be observed upon heating at 400 K (Figure 5 d and f) and as mentioned, are reversible upon cooling. On the neutron diffractogram, the Bragg peaks at 16 °, 16.6 ° and 17.8 ° merge into a broad peak upon heating at 400 K while the intensity of the strong peak at 19.5 ° decreases drastically. The peak at 24.2 ° disappears. Furthermore, the peaks at 19.5 ° is slightly shifted towards higher angles. On the X-ray diffractogram, the intensity of the peaks at low and high angle decrease. It is clear from these observations that we are observing a reversible transition for PTTP-TEG upon heating that is not observed in the other samples. To gain further insight in the changes of morphology upon heating, we perform inelastic neutron scattering (INS).

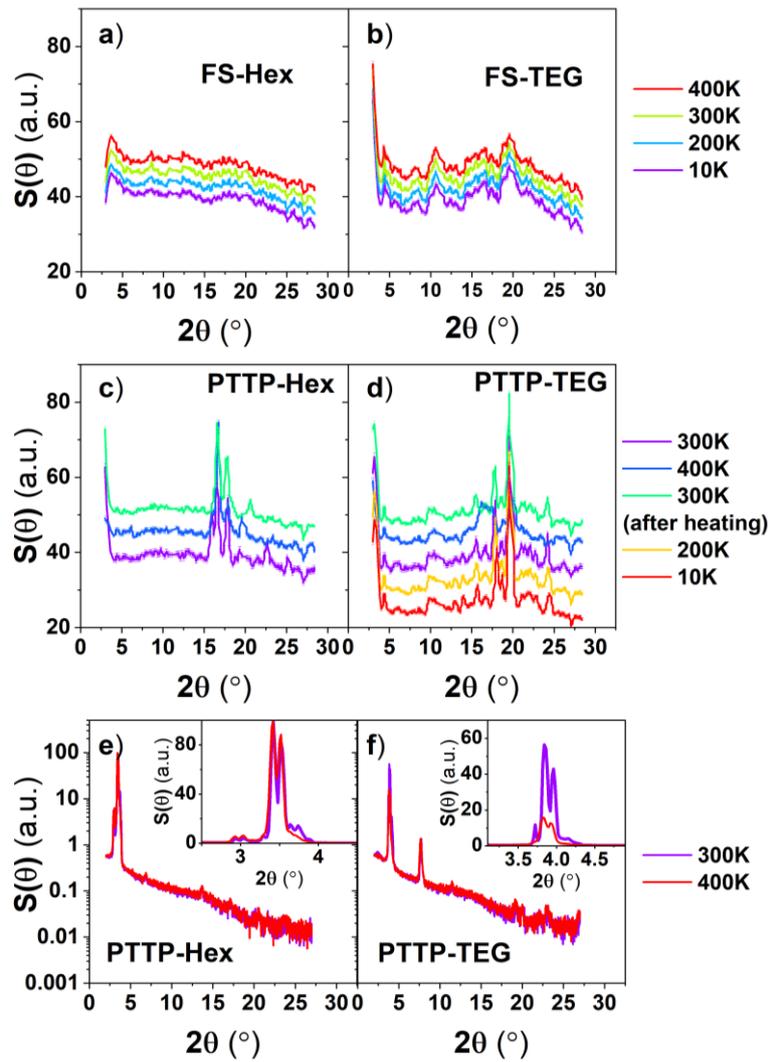

Figure 5. Temperature-dependent neutron diffractograms of (a) FS-Hex, (b) FS-TEG, (c) PTTP-Hex, (d) PTTP-TEG. Temperature-dependent specular X-ray diffractograms of (e) PTTP-Hex and (f) PTTP-TEG.

Figure 6 (a-b) shows the generalized density of state (GDOS)‡ of the set of samples at 300 K. The GDOS of the FS polymers are almost featureless in comparison with the PTTP small molecules pointing towards more disorder in the polymeric systems. This is in good agreement with the neutron diffractograms showing an increased crystallinity for the PTTP small molecules. The FSs exhibit two bands at about 15.5 meV and 30 meV. The bands appear more pronounced for FS-TEG and the Debye growth is slightly shifted to lower energy for FS-Hex, indicating a slight softening of the modes for FS-Hex with respect to FS-TEG. This points towards a higher disorder in FS-Hex with respect to FS-TEG as also observed by diffraction. PTTP-Hex exhibits a broad band at about 15 meV. For PTTP-TEG, this broad band is better resolved revealing a strong band at 12 meV and a weaker band at about 20 meV. An extra band appears for PTTP-TEG at about 30 meV. The Debye growth of PTTP-Hex is shifted to lower

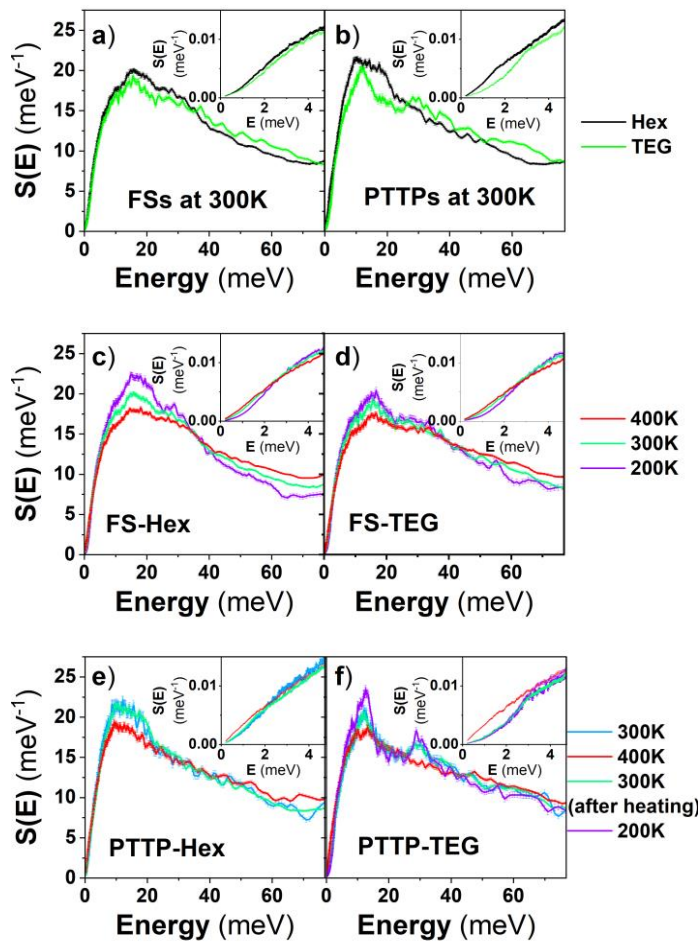

Figure 6. Generalized density of state (GDOS) of (a) FSs and (b) PTTPs at 300K after annealing. Temperature-dependent GDOS of (c) FS-Hex, (d) FS-TEG, (e) PTTP-Hex and (f) PTTP-TEG. The insets show the Debye growth. For the FSs, the samples were first measured at 400K, then at 300K and finally at 200K. For the PTTP molecules, the samples were first measured at 300K, then 400K, then 300K and finally 200K for PTTP-TEG

energy indicating a softening of the mode with respect to PTTP-TEG. Both the better resolution of the vibrational bands for PTTP-TEG and the clear softening of the modes for PTTP-Hex points towards an increased disorder for PTTP-Hex that could not be clearly seen in the diffractograms. The difference in Debye growth between PTTP-Hex and PTTP-TEG is more pronounced than between the FS polymers with a clear shift towards lower energy for PTTP-Hex. This highlights a general softening of the modes with the alkyl chains, leading to an increased disorder.

Upon cooling from 400 K to 200 K, the GDOS of both FS-Hex and FS-TEG becomes more resolved due to the decrease of the Debye-Waller effect and, the Debye growth shows a shift towards lower energies showing a slight hardening of the modes (Figure 6 c-d). Upon cooling from 300 K to 200 K, the GDOS of PTTP-TEG becomes also more resolved and the GDOS becomes dominated by the band at 12 meV and 30 meV. Unlike for the FSs, no changes in the Debye growth is observed upon cooling to 200 K (Figure 6 f). The PTTPs behave differently compared to the FSs upon heating at 400 K (Figure 6 e-f). The broad band at about 15 meV disappears for PTTP-Hex and the bands at 12 meV and 30 meV disappear for PTTP-TEG. A small shift towards lower energies in the Debye growth is observed for PTTP-Hex and a pronounced shift is observed for PTTP-TEG. The modes of the PTTPs soften upon heating, especially for PTTP-TEG. The GDOS of PTTP-TEG resembles at 400 K the featureless GDOS of PTTP-Hex and both GDOS exhibit similar Debye growth. Although the diffractograms of PTTP-Hex and PTTP-TEG are different at 400K, the GDOS of the two materials are similar. This indicates that the morphology of the two materials seem to be similar while the microstructure is different. We hypothesise based on diffraction, POM and INS measurements that the transition between 300 and 400 K for PTTP-TEG is due to the TEG chain melting, impacting both the crystalline structure and the disorder of the material. To capture the change in chain dynamics, we further performed quasi-elastic neutron scattering (QENS) using the IN6 spectrometer offering an energy window suitable to capture chain motions for the studied systems.

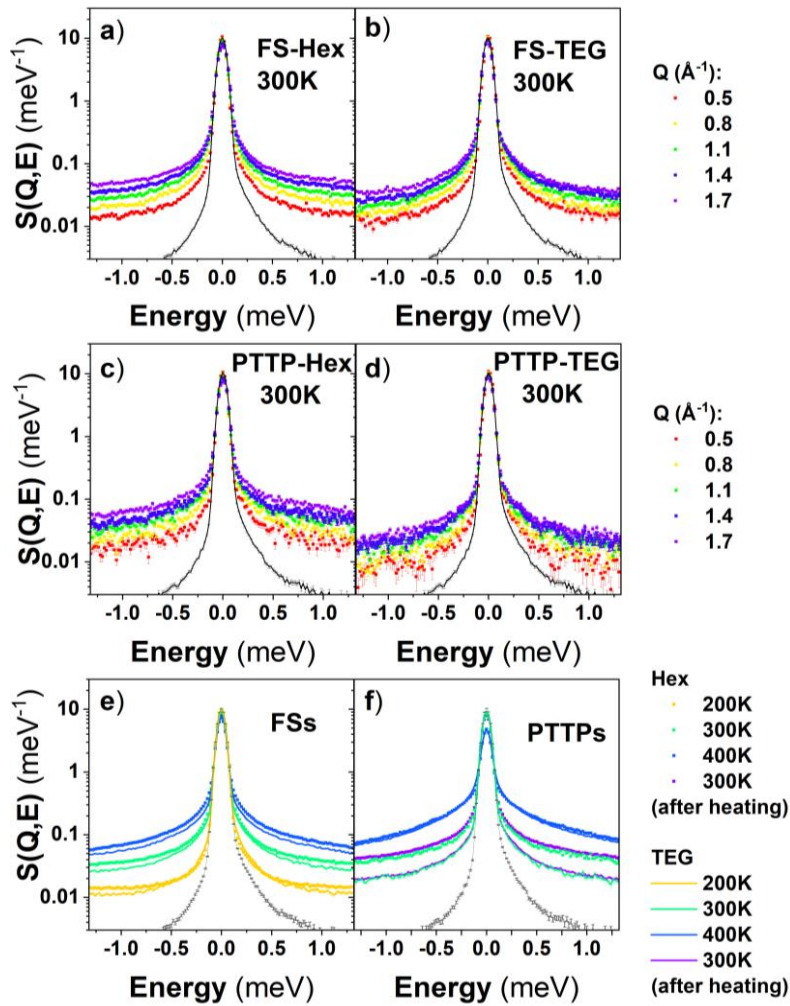

Figure 7. Q-dependent QENS spectra of (a) FS-Hex, (b) FS-TEG, (c) PTTP-Hex and (d) PTTP-TEG at 300 K after annealing. Temperature-dependent Q-averaged QENS spectra of (e) FSs and (f) PTTPs. The black spectra are the QENS spectra of FS-Hex for the FS samples and PTTP-TEG for the PTTP samples measured at 10 K and thus, represents the instrumental resolution.

Figure 7 (a-d) shows the Q-dependent QENS spectra of all the studied samples at 300 K. In all cases, a clear Q-dependence is observed. It appears that the QENS spectra are overall narrower for FS-TEG and PTTP-TEG than FS-Hex and PTTP-Hex, respectively, but with a similar Q-dependence. This is better shown by averaging the QENS spectra over the Q range (Figure 6 e-f). The FS-TEG exhibit slightly narrower QENS spectra for all the measured temperatures (Figure 6 e). This can be due to several factors: (i) the difference in numbers of hydrogens, (ii) the results of FS-TEG being slightly more ordered than FS-Hex, as observed on the diffractograms and GDOS, leading to an extra elastic contribution or (iii) a slower dynamics of the TEG chains with respect to the Hex chains. The QENS spectrum of PTTP-TEG at 300 K before and after annealing are superimposed (Figure 6 e) while the QENS spectrum of PTTP-Hex at 300K, before annealing, is slightly narrower than after annealing (Figure 6 f). This can be due to the observed difference in the diffractograms of PTTP-Hex before and after annealing. The polymorphs that melt upon annealing and do not crystallise upon cooling exhibit either more frustrated motions of the chains or exhibit an extra elastic contribution. The fact that the GDOS seems unaffected would point towards frustration.

Upon heating, the QENS spectra of all the samples broaden as expected since the chains' motions become faster with temperature, once activated. The QENS spectra of FS-TEG remains slightly narrower for all the measured temperatures with respect to the QENS spectra of FS-Hex. Interestingly, the QENS spectrum of PTTP-TEG is significantly narrower at 300 K than the PTTP-Hex while the QENS spectrum of PTTP-TEG resembles the QENS spectrum of PTTP-Hex at 400 K (Figure 6 f) with a slight narrowing similar to what is observed for the FSs. The phase transition observed for PTTP-TEG in both diffractograms and GDOS is reflected in the QENS and clearly involves fastening of the TEG chain motions.

## Experimental

The two FS polymers (Hex and TEG) used in the present neutron measurements were the same already synthesised and studied in reference 23.

Both small PTTP molecules (Hex and TEG) were synthesised according to previously published procedures.[20] X-ray powder diffraction was carried out on a PANalytical X'Pert Pro diffractometer using Ni-filtered Cu Kα radiation, setup in grazing incidence configuration. Thin films were drop cast onto boron doped single side polished silicon substrates of <100> orientation, purchased form PI-KEM, from 10 mg ml-1 polymer solutions in 1,2-dichlorobenzene. Films were covered with a glass petri dish and allowed to dry overnight in air. DSC was carried out on a Perkin Elmer DSC4000. In all calorimetry diagrams, endothermic processes are up and exothermic processes are down. General procedure for differential scanning calorimetry: 1) Hold for 1.0 min at 0.00°C - Heat from 0.00°C to 300.00°C at 10.00°C/min; 2) Hold for 1.0 min at 300.00°C - Cool from 300.00°C to 0.00°C at 10.00°C/min;

3)Hold for 1.0 min at 0.00°C - Heat from 0.00°C to 300.00°C at 10.00°C/min; 4) Hold for 1.0 min at 300.00°C - Cool from 300.00°C to 0.00°C at 10.00°C/min; 5) Hold for 1.0 min at 0.00°C - Heat from 0.00°C to 300.00°C at 10.00°C/min; 6) Hold for 1.0 min at 300.00°C - Cool from 300.00°C to 0.00°C at 10.00°C/min; 7) Hold for 1.0 min at 0.00°C.

Single crystal X-ray diffraction data for PTTP-TEG were obtained on a 0.03 mm × 0.04 mm × 0.05 mm block using Cu-Kα radiation on a Rigaku 007HF diffractometer equipped with Varimax confocal mirrors and an AFC11 goniometer and HyPix 6000HE detector at the National Crystallography Service at the University of Southampton, UK. The structure was solved by Patterson methods using the program SHELXT[31] and refined using SHELXL[32] in space group $P2_1/c$ with $a$ = 22.9690(10) Å, $b$ = 8.1561(3) Å, $c$ = 8.6458(3) Å, $\beta$ = 100.575(4)° and volume = 1592.17(11) Å$^3$. The final anisotropic full-matrix least-squares refinement on F$^2$ with 209 variables converged at R$_1$ = 4.88%, for the observed data and wR$_2$ = 13.88% for all data. CCDC 2052663 contains the supplementary crystallographic data for this paper. These data can be obtained free of charge from The Cambridge Crystallographic Data Centre via www.ccdc.cam.ac.uk/data_request/cif. Polarised optical microscope (POM) measurements were obtained using a Linkam LTS 350 hot stage and controller and an XPL3202 polarising microscope with digital camera measurements carried out in transmission using a 5 micron cell.

The neutron scattering measurements were performed using the direct geometry, cold neutron, time-of-flight, time-focusing spectrometer IN6, at the Institut Laue-Langevin (ILL, Grenoble, France). An optimized sample thickness of 0.2 mm was considered, relevant to the minimization of effects like multiple scattering and absorption. The temperature-dependent QENS and GDOS$^\ddagger$ spectra were collected using an incident neutron wavelength of 5.12 Å (Ei ≈ 3.12 meV), offering an optimal energy resolution at the elastic line of ~ 0.07 meV. Data were reduced, treated and analysed in a similar way as was done in our closely related work on conjugated polymers.[33]

## Conclusions

We studied the impact of the substitution of alkyl (n-hexyl) chains by triethylene glycol (TEG) chains on two families of materials, fluorene-co-dibenzothiophene sulfone (FS) polymer with the chains perpendicular to the backbone and phenylene-bithiophene-phenylene (PTTP) small molecule with the chains at the peripheral end-group positions.

A correlated interpretation of diffraction patterns from X-Ray and neutron measurements allowed to study the long-range order of the materials. We used complementarily INS to probe the associated morphologies. Both FS polymers are largely amorphous with FS-Hex slightly softer than FS-TEG. Unlike the FS polymers, the PTTP small molecules are crystalline. PTTP-Hex also appears softer than PTTP-TEG. This softening is reflected in the QENS measurements by broader QENS spectra. We highlight a general softening of the modes induced by the presence of alkyl chains and leading to an increased dynamical disorder of PTTP-Hex and FS-Hex.

Temperature-dependent measurements, including calorimetry measurements, evidence an extra exothermic transition close to room temperature and its associated endothermic transition at about 95 ˚C for PTTP-TEG in comparison to PTTP-Hex. The transition is shown to be reversible. At 400K, the diffractograms of PTTP-Hex and PTTP-TEG are different while the GDOS are similar. The microstructures of the samples at 400 K exhibit differences, while their morphologies resemble each other. The reversible transition observed for PTTP-TEG structurally is clearly reflected in the QENS measurement. We are likely capturing the reorientation of the TEG chains within the energy/time window of the instrument. The motions of the chains of PTTP-TEG are experiencing a change in dynamics upon heating not observed for PTTP-Hex neither for the FS polymers. Furthermore, POM shows no noticeable differences in birefringence upon heating for both materials. Therefore, we assign the observed structural transition of PTTP-TEG upon heating to the melting of the TEG chains.

Alkyl chains have been substituted by TEG chains for aqueous electrolyte-gated transistors and photocatalytic water-splitting to increase the interfacial area by allowing water uptake. Swelling is well-known to occur more readily in less ordered materials but good charge transport needs to be maintained. TEG chains promote overall order in the FS polymer while the TEG chains disturb the planarity of PTTP and crystallise below room temperature. Therefore, by promoting order and increasing polarity in FS, TEG chains could potentially be beneficial for water splitting applications as the TEG chains promote swelling and help maintaining the charge transport. The picture is more complex for PTTP as the TEG chains disturb the planarity of the molecules and the herringbone configuration is modified upon the substitution of the Hex chains, which can affect the charge transport. The crystallisation of the TEG chains for PTTP below room temperature can potentially lead to a stronger resilience of the charge transport upon swelling of the materials in OECTs but can also block the ion transport in these molecules. Thus, we hypothesise that if the ion transport is limiting in PTTP-based OECTs longer or branched TEG chains may limit the observed chain ordering below room temperature in PTTP upon substitution and lead to better materials.

## Author Contributions

A.A.Y.G. and M.Z. conceived and developed the neutron-based project, performed the neutron measurements, treated, analysed and interpreted the neutron data. Z.S.P. and C.B.N. synthesized the PTTP materials and performed the associated characterisations with T.K. and I.A. D.J.W. provided the FS-TEG Material and R.S.S. provided the FS-Hex material. A.A.Y.G. and M.Z. wrote the manuscript with contribution from Z.S.P., I.A. and C.B.N.

## Conflicts of interest

There are no conflicts to declare.

## Acknowledgements

The ILL is thanked for providing beam time on the IN6 spectrometer. Prof. Andy Cooper at the University of Liverpool is thanked for discussion regarding the FS materials. A. A. Y. G. acknowledges EPSRC for the award of an EPSRC Postdoctoral Fellowship (EP/P00928X/1). Z.S.P. and C.B.N. acknowledge support from the Academy of Medical Sciences & Wellcome Trust (SBF002/1158) and the Materials Research Institute.

## Notes and references